\def\year{2018}\relax
\documentclass[letterpaper]{article} 
\usepackage{aaai18}  
\usepackage{times}  
\usepackage{helvet}  
\usepackage{courier}  
\usepackage{url}  
\usepackage{graphicx}  
\begin{document}
%
\title{Using Pupil Diameter to Measure Cognitive Load}
\author{Georgios Minadakis and Katrin Lohan\\
Heriot-Watt University\\
MACS Department\\
Edinburgh\\
}
\maketitle
\begin{abstract}
In this paper, we will present a method for measuring cognitive load and online real time feedback using the Tobii Pro 2 eye-tracking glasses. The system is envisaged to be capable of estimating high cognitive load states and situations, and adjust human machine interfaces to the user's needs. 
The system is using well-known metrics such as average pupillary size over time. Our system can provide cognitive load feedback at 17-18 Hz. 
We will elaborated on our results of a HRI study using this tool to show it's functionality.
\end{abstract}
\vspace{-0.5cm}
\section{Introduction}
\noindent In human-machine interaction information delivery and interface design has been researched extensively.
User centered design is trying to focus on the users needs and iteratively develops better or more appropriate interfaces. This concept is challenging and long-term through it's iterative nature. Interfaces need to be tailored to users as well as adapted to high stakes, hazardous tasks.
In this work we present a system that will use measurements of cognitive load through the Tobii Pro 2 eyetracking glasses, which allows to measure pupil diameter in real-time. 
\\
For decades now, it has been established that changes in one's eyes' pupils diameter is an indicator of cognitive activity. In the beginning of the previous century the German neurologist Bumke had already recognized that every intellectual or physical activity translates into pupil enlargement \cite{article}. 
\\
Pupillometry is the measurement of pupil size and reactivity, is a key part of the clinical neurological exam for patients and evaluates the pupils of patients with the focus on the pupil size.
Pupillary responses can reflect activation of the brain allocated to cognitive tasks. Greater pupil dilation is associated with increased processing in the brain \cite{GRANHOLM20041}.
\\
Hess and Polt achieved a major milestone for \textit{pupillometry} by discovering that showing semi nude photos of adults to subjects of the opposite sex would cause their pupils to dilate twenty percent on average \cite{HessEckhardH.1960Psar}. This study provided evidence that emotional stimulation causes enlargement of pupil diameter. This notion was later expanded upon to include more cognitive processes, such as memory and problem solving. Beatty and Kahneman showed that storing an increasing number of digits in one's memory would cause pupillary dilation \cite{BeattyJackson1966Pcit}, while it was also shown by Hess and Polt that pupil size corresponds with the difficulty of a cognitive task \cite{HessEH1964PSiR}.
\begin{figure}
\includegraphics[width=0.45\textwidth]{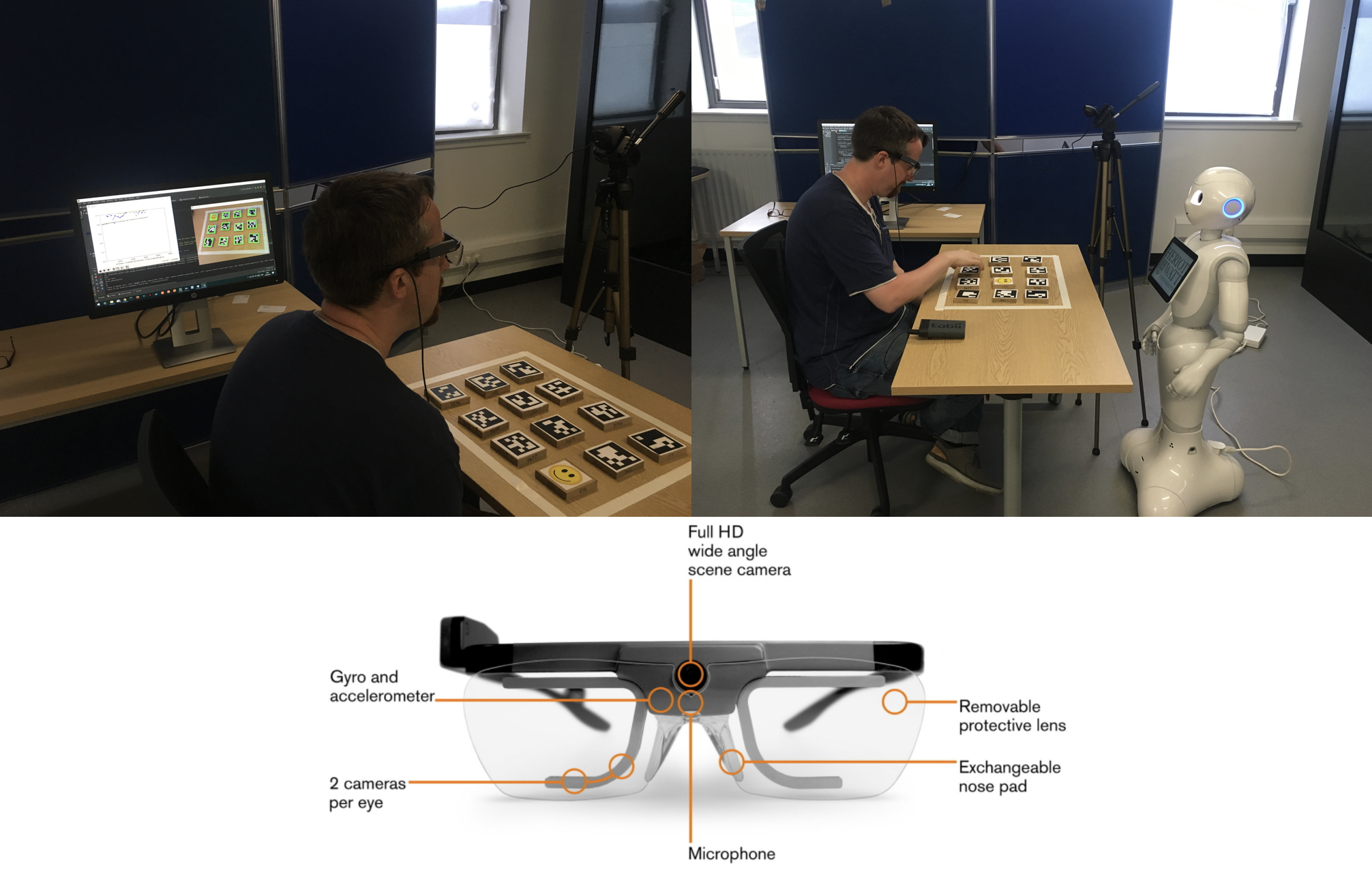}
\caption{System setup: The tobii glasses 2 are given to each of our participants our software records their pupil diameter during the interaction game with the robot. When the cognitive load is perceived as high this can be observed by the experiment as it is display on a screen during the interaction. }
\label{pic3}
\vspace{-0.8cm}
\end{figure}
More recently, Just and Carpenter \cite{JustMarcelAdam1993TIDo} showcased that pupil responses can be an indicator of the effort to comprehend and process information. They conducted an experiment where participants were given two sentences of different complexities to read while they would measure their pupil diameters. Pupillary dilation was larger while readers processed the sentences that were deemed to be more complicated and more subtle while they read the simpler one \cite{JustMarcelAdam1993TIDo}. We believe these findings make the connection between \textit{pupillometry} and cognitive load theory clear, as they demonstrate that changes in the properties of an element to be processed (e.g. changing the complexity of a sentence affects the amount of intrinsic load it will impose to the reader), cause different pupillary responses.
While cognitive load can be affected by a large number of factors, we believe that pupillometry offers a responsive signal that can potentially provide approximate real-time feedback of the users arousal and potentially their cognitive load. Techniques in pupillometry have been successfully employed to measure load in past studies like (\cite{palinko2010estimating,klingner2010measuring},etc.).
Pupillometry could be a powerful tool to measure cognitive load but could be affected by other confounding factors.
\vspace{-0.6cm}
\section{Online System}
Our system connects the Tobii Pro 2 eye-tracking glasses using the Tobii Pro Glasses 2 python controller \cite{TobiiProGlasses2} library over wifi with the computer (see figure\ref{pic3}). The new system provides information on the cognitive load of the participant at 17 fps. After a very simple calibration phase that is taking advantage of the pupillar light reflex \cite{kun2012exploring}, the system is able to track the users cognitive load, based on an empirically set threshold. Our system provided a running average $\zeta$, of pupil diameter:
\vspace{-0.2cm}
$$\zeta(d)=\frac{\sum_{t=1}^N{d_t}}{N}$$
where $d_t$ is the current pupil diameter and $N$ the number of frames.  
It further provides a windowed average:
\vspace{-0.1cm}
$$\zeta(d)=\frac{\sum_{t=1}^{15}{d_t}}{15}$$
where the average is calculated based on the past 15 frames only.
Furthermore, our system provides a running peek estimation where we assume that, when the size of the pupil is larger than  $70\%$ of the maximum the cognitive load is high.
\vspace{-0.2cm}
\section{Human Robot Interactions (HRI) study}
We run a HRI study with our system and two robots the Pepper robot and the Husky robot. Participants were invited to play a pairs matching game with robots. The robot would assist the participants, giving clues about where the matching pair would be and comment on the participants performance. The participants received more points when asking the robot for clues.   
\vspace{-0.6cm}
\section{Participants}
We had 31 participants between 25-46 years with a mean age of 30.9 interacting with either of the  robots in either of 2 conditions (high or low error rate). We received 8 participants to interact with the Pepper at a low error rate, 9 participants to interact with the Pepper at a high error rate, 5 participants to interact with the Husky at a low error rate and 9 participants to interact with the Husky at a high error rate. 
\vspace{-0.2cm}
\section{Evaluation}
\begin{figure}[t]
\includegraphics[width=0.5\textwidth]{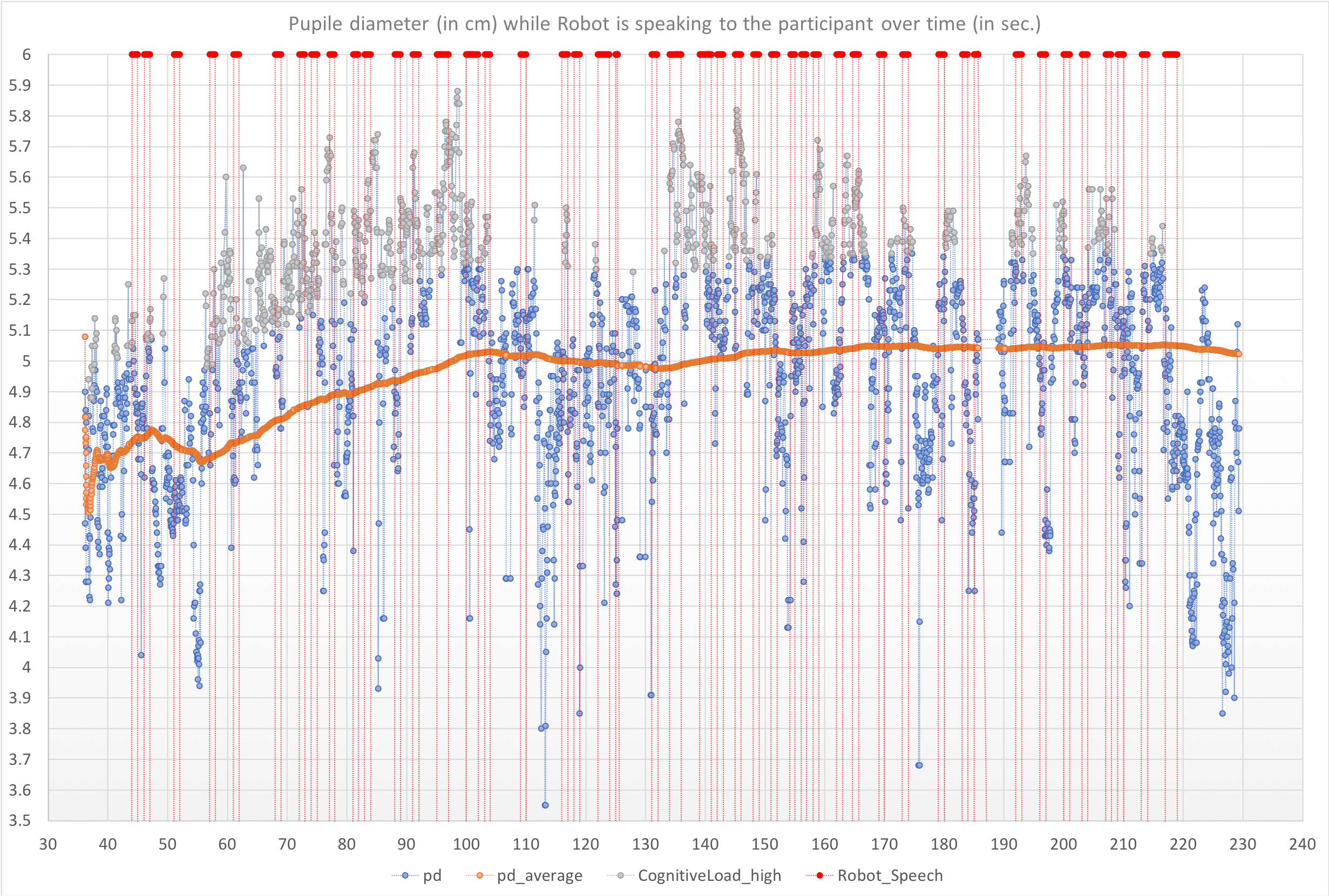}
\caption{The pupil diameter of the participant whilst the robot is talking to the participant.}
\label{pic2}
\vspace{-0.5cm}
\end{figure}

In our experiment, we have participants discover 6 matching pairs from overturned cards and the robot would support them with clues, which were either perfectly correct in the low error rate conditions or not always correct in the high error rate condition. We run a Pearson correlation between amount of questions for support participants asked and the number of peaks (maximal cognitive load) and found a negative correlation between the two variables, r = -.328, p = .041. This suggests that if the participants decided to search without help for the matching pairs they were cognitively more loaded than if they ask for the matching pair from the robot. 
Furthermore, we found a negative correlation between number of peaks and session duration , r = -.578, p $<=$ .001. This suggests if participants did not rely on the help from the robot, but therefore needed more attempts to find all 6 pairs, they had a higher cognitive load over the session. This is also supported by the fact that we found a negative correlation between number of tries participants needed to find all pairs and the session duration, r = -.680, p $<=$ .001. 
Finally, we looked into how participants reacted to the robots greeting them. Here we found a negative correlation between the number of peaks (maximal cognitive load) and the condition they where in, r = -.535, p $<=$ .001. 
For a deeper investigation of this we looked into one participant interacting with the Pepper robot in the low error rate condition and found that the cognitive load might be preemted by the robot talking to the participant (see Figure \ref{pic2}).

\vspace{-0.3cm}
\section{Discussion and Conclusion}
We can see that there are correlations suggesting that when participants decided to search without help for the matching pair they are cognitively more loaded than if they ask for the matching pair.
This might be related to the condition they are in. So if the robot was in a low error rate condition they were asking more for help than if they were in a high error rate condition. We have not yet investigated this aspect of the data. Furthermore, there might be a correlation with age of the participants and the cognitive load that we have not investigated yet. Using pupil diameter as a measure for cognitive load might depend on age, as the reaction time of the pupil changes during aging. 
Nevertheless, we found relative stable results in the detection of cognitive load in the pupil diameter and we believe therefore it is a good real-time measure for cognitive load. Other measures of cognitive load are of interest for us as well, like verbal features (e.g pitch, volume or velocity) which have recently investigated here \cite{Lopesetal2018}. 
Future work seeks to collect more data to establish these results presented are robust with different participants and that we can generate novel task. So far, our system shows promising way of detecting cognitive load in HRI but further evaluation and data collection are needed. 
\vspace{-0.6cm}
\section{Acknowledgements}
The authors would like to acknowledge the support of the EPSRC IAA 455791 along with ORCA Hub~EPSRC (EP/R026173/1, 2017-2021) and consortium partners.
\bibliographystyle{aaai}
\bibliography{pupilar}
\end{document}